# Neutrally floating objects of density ½ in three dimensions

Péter L. Várkonyi
Budapest University of Technology and Economics, Hungary
vpeter@mit.bme.hu

*Abstract*
*This paper is concerned with the Floating Body Problem of S. Ulam: the existence of objects other than the sphere, which can float in a liquid in any orientation. Despite recent results of F. Wegner pointing towards an affirmative answer, a full proof of their existence is still unavailable. For objects with cylindrical symmetry and density ½, the conditions of neutral floating are formulated as an initial value problem, for which a unique solution is predicted in certain cases by a suitable generalization of the Picard-Lindelöf theorem. Numerical integration of the initial value problem provides a rich variety of neutrally floating shapes.*

## 1. Introduction

The equilibria of floating objects subject to gravity and buoyancy forces have surprising properties. The stable equilibria of symmetrical objects are often asymmetrical (Gilbert, 1991; Erdős et al, 1992a,b; Bass, 1980; Nye and Potter, 1980). Alternatively, the set of equilibrium configurations may have symmetries exceeding the degree of the object's symmetry. An interesting question about floating objects – often referred to as Floating Body Problem – was proposed over seventy years ago by Stanislav Ulam as Problem 19 of the Scottish Book (Mauldin, 1981): „are spheres are the only bodies that can float (without turning) in any orientation?" A simpler two-dimensional version of this problem, also credited to Ulam, concerns the existence of non-circular logs with horizontal axis, which can float in every orientation. There are simple nontrivial solutions among disconnected bodies in two dimensions or shapes containing holes in 3 dimensions (Wegner, 2008). To exclude such solutions, both questions are commonly restricted to star-shaped bodies. In this paper, we require solutions to be simple with respect to $\rho=1/2$ according to

*Definition 1: a body is simple with respect to a 'density parameter' $\rho$ if every planar cut dividing its volume in ratio $\rho:1-\rho$ forms a simply connected set.*

Being simple and being star-shaped are closely related and both classes include convex objects. For star-shaped bodies, the planar problem was solved long ago by Auerbach (1938) for density $\rho=½$ relative to the liquid, and much more recently by Wegner (2003) for other densities. In both cases, many nontrivial neutrally floating objects have been identified. In three dimensions, there are no solutions in the limit $\rho \to 0$ or 1 (Montejano, 1974); and no solutions among star-shaped objects with central symmetry (other than the sphere) for density $\rho=½$ (Schneider 1970, Falconer, 1983). Nevertheless, F. Wegner has proposed a perturbation expansion scheme starting from the sphere for objects with central symmetry and $\rho \neq ½$ (Wegner, 2008), as well as for bodies with arbitrary shape and $\rho= ½$ (Wegner, 2009). His results point towards the existence of many nontrivial solutions in these wider classes of shapes, even though the proofs are incomplete in that the convergence of the perturbation series has not been examined. Furthermore, no attempt to construct actual solutions of the problem has been reported.

We take a different approach to construct three-dimensional, neutrally floating objects of density $\rho=1/2$ with cylindrical symmetry. Our method is an adaptation of Auerbach (1938) to the three-dimensional problem. After reviewing the geometric conditions of neutral floating in Section 2.1-2.2, these are transformed into a non-standard integro-differential equation with given initial conditions (ie. an initial value problem) for the generating curve of the object (Section 2.3-2.4) using fractional order derivatives. It is shown in Section 3 that sufficiently small perturbations of the sphere yield physically meaningful nontrivial solutions of the problem and examples are constructed by integrating the equations numerically. The paper is closed by a short discussion of related problems.

## 2. Equations of neutral floating bodies

### 2.1. Geometric criteria of neutral floating

By the principle of Archimedes, a body of density ρ floats in a liquid of density 1 in such way that a fraction ρ of the object's volume is immersed in the liquid. A configuration satisfying Archimedes' principle is an equilibrium iff the centroid of the object (G) is exactly above the centroid of the immersed portion. The equilibrium is neutral, if after small rotations (with the preservation of Archimedes' principle), the centroid of the immersed part remains on a sphere centered at G, yielding constant potential energy. Our goal is to design objects, for which every configuration satisfying Archimedes' principle is a neutral equilibrium, i.e. for which the centroids of the immersed parts for every possible configuration form a sphere of arbitrary radius $r$ centered at G.

Any plane that divides the object's volume in ratio ρ:1-ρ is called a *water plane* ($W$) and the intersection of the object with any water plane $W$ as *water section* or $W^*$. We consider two water planes infinitesimally close to each other. The transformation mapping one ($W_1$) to the other ($W_2$) is a rotation by an infinitesimal angle $\alpha_{12}$ about a line $l_{12}$. The water planes and sections have two remarkable properties described below. For a more detailed description, the reader is advised to consult Gilbert (1991), Wegner (2007) or references therein.
1) The conservation of the immersed volume implies that $l_{12}$ goes through the centroid of $W_1^*$. *As a consequence, the union of the centroids of water sections is a 'water envelope' E (more precisely a wavefront possibly containing singularities) such that every water plane is tangential to E.*
2) If $W_1$ corresponds to a neutral equilibrium, then the distance between the centroids of the immersed volumes ($G_1$ and $G_2$) corresponding to the two water planes is $|G_1G_2| = r\alpha_{12}$. The same distance can also be expressed as $|G_1G_2| = \alpha_{12} I_{12} (\rho V)^{-1}$ where $V$ is the volume of the object and $I_{12}$ is the moment of inertia of $W_1^*$ about the axis $l_{12}$. Thus neutrally floating bodies are characterized by the additional property that, *the moment of inertia of any water section, about any axis going through its centroid is constant I.*

Property (ii) is necessary but not sufficient characterization of a neutral equilibrium since the sphere formed by the centroids is not necessarily centered at G. However, for objects of density ½, the centroid G is exactly halfway between the centroid of the submerged part ($G_1$) and centroid of the rest of the object ($G_1'$) Furthermore, $G_1$ and $G_1'$ are opposite points of the above mentioned sphere. Hence, the sphere is centered at G, i.e. the requirement of constant inertia is necessary and sufficient.

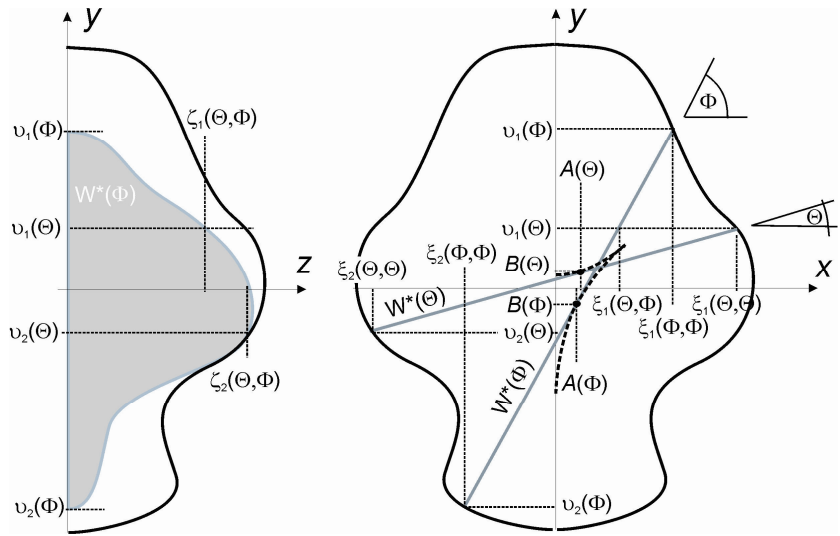

**Figure 1**: Projections of the object to the *x-y* (right panel) and *y-z* (left panel) plane. Thick solid lines denote its contour, *y* is the symmetry axis. The dashed curve in the right panel is the contour of the water envelope. W*(Θ) and W*(Φ) are two water sections, which are parallel to the *z* axis. W(Φ) is also shown in the left panel by grey shading. For further notations, see the main text.

### 2.2. Integral equations of neutral floating

After introducing the notations of the paper, we develop equations corresponding to the conditions of neutral floating. We restrict our attention to objects, which are invariant to arbitrary rotation about axis *y* of a Cartesian coordinate system *x-y-z*. Due to the rotational symmetry, it is enough to consider water planes and sections parallel to the *z* axis. Let $W^*(\Phi)$ denote one such water section, which is at angle $0 \le \Phi \le \pi/2$ to the *xz* plane (Fig. 1). According to property 1) of Section 2.1, the centroid of $W^*(\Phi)$ belongs to the contour of the rotation-symmetric water envelope *E*. The *x* and *y* coordinates of the centroid of $W^*(\Phi)$ are $A(\Phi)$ and $B(\Phi)$. The intersection of $W^*(\Phi)$ with the *x-y* plane is a line section. Let $\upsilon_1(\Phi)$ and $\upsilon_2(\Phi)$ denote the signed distances of its endpoints from the *x-z* plane (such that $\upsilon_2$ is usually negative). The functions $A$, $B$, $\upsilon_1$ and $\upsilon_2$ determine the object's shape uniquely.

Let $0 \le \Theta \le \Phi$. The intersection of $W^*(\Phi)$ with the plane $y = y_j(\Theta)$ is a line section parallel to the *z* axis. The *x* coordinate of this section is

$$\xi_j(\Theta, \Phi, \upsilon_j(\Theta)) = A(\Phi) + (\upsilon_j(\Theta) - B(\Phi))\cot\Phi .\tag{1}$$

and the half-length of the same line section is by Pythagoras' theorem:

$$\zeta_j(\Theta, \Theta, \upsilon_j(\Theta)) = \sqrt{\xi_j^2(\Theta, \Theta, \upsilon_j(\Theta)) - \xi_j^2(\Theta, \Phi, \upsilon_j(\Theta))} \tag{2}$$

We introduce new variables $\alpha = \sin\Theta$, $\phi = \sin\Phi$ and functions $a(\alpha) = A(\Theta)$; $b(\alpha) = B(\Theta)$; $Y_j(\alpha) = \upsilon_j(\Theta)$; $X_j(\alpha, \phi, Y_j(\alpha)) = \xi_j(\Theta, \Phi, \upsilon_j(\Theta))$, which will lead to more convenient equations later. Then, (1), (2) become

$$X_j(\alpha, \phi, Y_j(\alpha)) = a(\phi) + (Y_j(\alpha) - b(\phi))\phi^{-1}(1 - \phi^2)^{1/2} = (\widetilde{a}(\phi) + (Y_j(\alpha) - b(\phi))\phi^{-1})(1 - \phi^2)^{1/2} \tag{3}$$

$$Z_j(\alpha, \phi, Y_j(\alpha)) = \sqrt{X_j^2(\alpha, \alpha, Y_j(\alpha)) - X_j^2(\alpha, \phi, Y_j(\alpha))} \tag{4}$$

where

$$\widetilde{a}(\alpha) \stackrel{def}{=} a(\alpha)(1 - \alpha^2)^{-1/2} . \tag{5}$$

Now we are ready to transform the two criteria of Section 2.1 into equations. By the definition of geometric centroids, the coordinate $b(\phi)$ of the centroid of $W^*(\arcsin\phi)$ satisfies

$$\sum_{j=1}^{2}(-1)^i \int_0^\phi Z_j(\alpha, \phi Y_j(\alpha))(Y_j(\alpha) - b(\phi))Y_j'(\alpha)d\alpha = 0 \tag{6}$$

where prime means derivative. Property 2) of Section 2.1 applied to an axis parallel to *z* can be expressed as

$$\sum_{j=1}^{2}(-1)^{j+1} \int_0^\phi Z_j(\alpha, \phi Y_j(\alpha))(Y_j(\alpha) - b(\phi))^2 Y_j'(\alpha)d\alpha = I\phi^3 \tag{7}$$

Notice that the left side of the equation is the moment of inertia of a projection of $W^*(\arcsin\phi)$ to the *y-z* plane rather than of $W^*(\arcsin\phi)$ itself. This is compensated by the $\phi^3$ term on the right side.

Due to the rotational invariance of the object, equilibria are always neutral against an infinitesimal rotation about an axis normal to *z*. Thus, property 2) holds for such axes and need not be checked. In summary, if the water envelope is given, then (6),(7) are necessary and sufficient conditions of neutral floating.

### 2.3. Steps towards an initial value problem

Analogously to the solution of the planar problem by Auerbach (1938), we first choose a water envelope (see Section 3.3 for more details). Once the functions *a* and *b* have been established, the integral equations (6),(7) depend on values of the functions $Y_i(\alpha)$ over the interval $\alpha \in (0,\phi)$. This observation suggests a transformation of the equations into an initial value problem. (6) and (7) can be written in the general form

$$\sum_{j=1}^{2} \int_{0}^{\phi} g_{ij}(\alpha,\phi,Y_j(\alpha))Y_j'(\alpha)d\alpha = f_i(\phi) \qquad (8)$$

where $f_j$ are scalar functions and $g_{i,j}$ are scalar functionals; $i = 1$ for the first equation and 2 for the second. Differentiating (8) with respect to $\phi$ yields

$$\sum_{j=1}^{2} \int_{0}^{\phi} \frac{\partial g_{ij}(\alpha,\phi,Y_j(\alpha))}{\partial \phi} Y_j'(\alpha)d\alpha + g_{ij}(\phi,\phi,Y_j(\phi))Y_j'(\phi) = f_i'(\phi). \qquad (9)$$

If the two by two matrix composed of the elements $g_{ij}(\phi,\phi,Y_i(\phi))$ is nonsingular, then $Y_j'(\phi)$ can be expressed explicitly from the equations, yielding a first-order initial value problem for $Y_j(\phi)$. Nevertheless it might happen that all elements of the matrix are zero. In this case, the second derivative of (8) becomes

$$\sum_{j=1}^{2} \int_{0}^{\phi} \frac{\partial^2 g_{ij}(\alpha,\phi,Y_j(\alpha))}{\partial \phi^2} Y_j'(\alpha)d\alpha + \left.\frac{\partial g_{ij}(\alpha,\phi,Y_j(\alpha))}{\partial \phi}\right|_{\alpha=\phi} Y_j'(\phi) = f_i''(\phi) \qquad (10)$$

which is again a candidate for an initial value problem. If the $\partial g_{ij}()/\partial \phi$ terms also happen to be zero, additional derivation of the equations might be necessary. Unfortunately, this method fails for the specific function $g_{ij}$ of the problem of neutral floating, because $g_{ij}(\phi,\phi,Y_i(\phi))$ is identically zero whereas the first derivative $\partial g_{ij}/\partial \phi$ does not exists; specifically

$$\lim_{\alpha \to \phi}\left|\frac{\partial g_{ij}(\alpha,\phi,Y_j(\alpha))}{\partial \phi}\right| = \infty \qquad (11)$$

The diverging limit indicates that the second derivative is "too much", whereas the first derivative of (8) is not enough. This special property of $g_{ij}$ is a consequence of the square-root type singularity of the function $Z_j$ in (4) at $\alpha=\phi$, inherited by $g_{ij}$. The specific form of $Z_j$ simplies that *the fractional derivative of order 3/2* of $g_{ij}$ is finite and nonzero at $\alpha=\phi$; thus the 1.5[th] derivative of (8) leads to an initial value problem.

## 2.4. Calculation of the fractional derivative

Fractional derivatives are defined as integer order derivatives of a fractional integral of order less than 1 (Miller&Ross, 1993). Thus, the first step towards the 3/2[th] derivative is to take the semi-integral of (8). The definition of Riemann–Liouville differintegrals yields

$$\sum_{j=1}^{2} \int_{0}^{\chi} (\chi-\phi)^{-1/2} \cdot \int_{0}^{\phi} g_{ij}(\alpha,\phi,Y_j(\alpha))Y_j'(\alpha)d\alpha\,d\phi = \int_{0}^{\chi} (\chi-\phi)^{-1/2} \cdot f_i(\phi)d\phi \qquad (12)$$

Before proceeding with the main steps, the order of integration is changed on the left side of the equation and the functions $G_{ij}$ and $F_{ij}$ are introduced:

$$\sum_{j=1}^{2} \int_{0}^{\chi} Y_j'(\alpha) \cdot \underbrace{\int_{\alpha}^{\chi} (\chi-\phi)^{-1/2} g_{ij}(\alpha,\phi,Y_j(\alpha))d\phi}_{G_{ij}(\alpha,\chi,Y_i(\alpha))} d\alpha = \underbrace{\int_{0}^{\chi}(\chi-\phi)^{-1/2} \cdot f_i(\phi)d\phi}_{F_i(\chi)} \qquad (13)$$

We differentiate both sides with respect to $\chi$, using the Leibniz integral rule:

$$\sum_{j=1}^{2}\left(\int_{0}^{\chi} Y_j'(\alpha) \cdot \frac{\partial}{\partial \chi} G_{ij}(\alpha,\chi,Y_j(\alpha))d\alpha + Y_j'(\chi) \cdot \underbrace{G_{ij}(\chi,\chi,Y_j(\chi))}_{\text{zero}}\right) = F_i'(\chi) \qquad (14)$$

The term $G_{ij}(\chi,\chi,Y_j(\chi))$ equals zero (see (43) in Appendix A.3). Thus we have,

$$\sum_{j=1}^{2} \int_{0}^{\chi} Y_j'(\alpha) \cdot \frac{\partial}{\partial \chi} G_{ij}(\alpha,\chi,Y_j(\alpha))d\alpha = F_i'(\chi) \qquad (15)$$

Differentiating both sides once more yields

$$\sum_{j=1}^{2}\int_{0}^{\chi}Y_j'(\alpha)\cdot\frac{\partial^2}{\partial\chi^2}G_{ij}(\alpha,\chi,Y_j(\alpha))d\alpha + Y_j'(\chi)\cdot\frac{\partial}{\partial\chi}G_{ij}(\alpha,\chi,Y_j(\alpha))\bigg|_{\alpha=\chi} = F_i''(\chi). \tag{16}$$

The unknowns $Y_j'(\chi)$ can be expressed explicitly as

$$\mathbf{Y}'(\chi) = \mathbf{A}(\chi,\mathbf{Y}(\chi))^{-1}\left(\mathbf{F}''(\chi) - \int_0^\chi \mathbf{C}(\alpha,\chi,\mathbf{Y}(\alpha),\mathbf{Y}'(\alpha))d\alpha\cdot\begin{bmatrix}1\\1\end{bmatrix}\right) \tag{17}$$

where $\mathbf{Y}$, $\mathbf{Y}'$ and $\mathbf{F}''$ are column vectors composed of the functions $Y_j$, $Y_j'$ and $F_j''$; $\mathbf{A}$ and $\mathbf{C}$ are 2 by 2 matrices with elements

$$a_{ij}(\chi,Y_j(\chi)) = \frac{\partial}{\partial\chi}G_{ij}(\alpha,\chi,Y_j(\alpha))\bigg|_{\alpha=\chi} \tag{18}$$

$$c_{ij}(\alpha,\chi,Y_j(\chi),Y_j'(\chi)) = Y_j'(\alpha)\cdot\frac{\partial^2}{\partial\chi^2}G_{ij}(\alpha,\chi,Y_j(\alpha)) \tag{19}$$

The function $\mathbf{F}''(\chi)$ can be expressed in closed form. We do not need to use any other property of $\mathbf{F}''$ than its boundedness in the forthcoming analysis. At the same time, we need to examine $\mathbf{A}$ and $\mathbf{C}$ thoroughly to analyze the solutions of the initial value problem (17).

## 3. Solutions

### 3.1. The existence and uniqueness of solutions

Spheres of any radius $R$ are neutrally floating objects. They correspond to $Y_j(\phi)=R(-1)^{j+1}\phi$. We deduce implicitly that this function satisfies the initial value problem (17) for $a(\phi)=b(\phi)\equiv 0$ with initial conditions $Y_j(0)=b(0)$. We refer to the corresponding equations and solutions as well as elements of these equations as *trivial* equations, solutions, etc. The one-parameter set of trivial solutions share the same initial condition. This is explained by the degenerate behavior of the variable $\mathbf{Y}(\chi)$ at $\chi=0$. Therefore, we need a second initial condition $\mathbf{Y}'(0)=R\cdot[1\ -1]^T$. Two further things should be noticed:
- if $\chi\to 0^+$, all elements $\mathbf{A}$, $\mathbf{C}$ and $\mathbf{F}$ in (17) go to zero, which could be compensated by a rescaling of the equation.
- (17) has been obtained by replacing an equation of the form $u(\chi,\mathbf{Y}(\chi),...)=0$ by its $3/2^{\text{th}}$ derivative. The transformed equations admits false solutions for which $u(\chi,\mathbf{Y}(\chi),...)=constant\cdot\chi^{3/2}$ rather than 0. Nevertheless, the set of false solutions correspond to water sections of second order moment $I+constant\cdot\chi^{-3/2}$, which contradicts any initial condition of the form $Y_j(0)=constant$.

In this section, we want to examine the effect of minor perturbations of $a(\phi)$ and $b(\phi)$. To avoid difficulties at $\chi=0$, we require that there is an interval $(0,\alpha_1)$ of $\phi$, where $a(\phi)=b(\phi)=0$ and $Y_j(\phi)=R(-1)^{j+1}\phi$. It is demonstrated below that the initial value problem has a unique solution under this restriction.

*Lemma 1* states that any solution of the perturbed equations must be close to the trivial solution, without examining if such solution(s) exist or not. The questions of existence and uniqueness are answered by *Lemma* 2. The two lemmas are summarized in *Theorem 1*, the main result of the paper.

*Lemma 1: for any given scalar $0<\alpha_1$, there exist positive scalars k and $\varepsilon_0$, such that if*
*(i) $a(\alpha)=0$, $b(\alpha)=0$ and $Y_j(\alpha)=(-1)^{j+1}\alpha$ if $0\leq\alpha\leq\alpha_1$;*
*(ii) the absolute values of $\tilde{a}(\alpha)$, $b(\alpha)$, and of their derivatives up to third order exist and they are $<\varepsilon<\varepsilon_0$ for any $\alpha_1\leq\alpha\leq 1$*
*then any solution of equations (17) over the interval $\alpha_1<\alpha\leq 1$ satisfies*

$$\left|Y_j'(\alpha)-(-1)^{j+1}\right|\leq k\varepsilon e^{k\alpha} \tag{20}$$

*Proof of Lemma 1:*
We arrive to (20) via proof by contradiction. The initial section $0 \leq \alpha < \alpha_1$ of $Y_i(\alpha)$ satisfies (20) for any $k$ by point (i) of *Lemma 1*. Let us assume now that (20) is violated no matter how large $k$ is. Then there must exist a unique scalar $\alpha_1 < \chi(k) < 1$ for any $k$ such that (20) holds for $\alpha \leq \chi(k)$ and there is equality in (20) for $\alpha = \chi(k)$ and $j=1$ or 2. In this case we also have

$$\left| Y_j(\alpha) - (-1)^{j+1}\alpha \right| \leq \int_0^\alpha \left| Y_j'(\beta) - (-1)^{j+1} \right| d\beta < \varepsilon \int_0^\alpha k e^{k\beta} d\beta = \varepsilon \left( e^{k\alpha} - 1 \right) < \varepsilon e^{k\alpha} \quad if \quad \alpha \leq \chi(k) . \tag{21}$$

From this point, the argument $k$ of $\chi$ is dropped for brevity.

If $\varepsilon$ is small enough, then (20) and (21) imply that
1) each entry of $\mathbf{A}(\chi, \mathbf{Y}(\chi))$ is within a neighborhood of radius $\ast \cdot \varepsilon e^{k\chi}$ of its trivial value, and the trivial value is bounded; $\ast$ represents some finite positive scalar, which is independent of $k$. For the proof, see Appendix A.1. Furthermore, $\mathbf{A}(\chi)$ is non-singular, i.e. $|\det \mathbf{A}|$ has a positive lower bound (see Appendix A.2). The two result imply that $\mathbf{A}^{-1}(\chi, \mathbf{Y}(\chi))$ is also within a neighborhood of radius $\ast \cdot \varepsilon e^{k\chi}$ of its bounded trivial value.
2) It is demonstrated in Appendix A.3 that if $k > 1$, then the second derivative of $G_{ij}$ is bounded

$$\left| \frac{\partial^2}{\partial \chi^2} G_{ij}(\alpha, \chi, Y_j(\alpha)) \right| < DDG_{ij}^{(\max)} . \tag{22}$$

and it is within distance $\ast \cdot \varepsilon e^{k\alpha}$ of its trivial value

$$\frac{\partial^2}{\partial \chi^2} G_{ij}(\alpha, \chi, Y_j(\alpha)) \in DDG_{ij}^{(0)}(\alpha, \chi) \pm \ast \cdot \varepsilon e^{k\alpha} . \tag{23}$$

By plugging (20), (22) and (23) into (19) one obtains
$$c_{ij} \in \left( (-1)^{i+1} \pm k \varepsilon e^{k\alpha} \right) \left( DDG_{ij}^{(0)}(\alpha, \chi) \pm \ast \varepsilon e^{k\alpha} \right) \in .... \tag{24}$$

$$\underbrace{(-1)^{i+1} DDG_{ij}^{(0)}(\alpha, \chi, Y_i)}_{\text{bounded trivial value}} \pm \underbrace{\left( DDG_{ij}^{(\max)} + \ast \cdot \underbrace{k^{-1}}_{<1} \right) \cdot \varepsilon k e^{k\alpha}}_{\ast \varepsilon k \exp(k\alpha) \text{ deviation}} \in$$

Hence, we conclude that each entry of $\mathbf{C}(\alpha, \chi, \mathbf{Y}(), \mathbf{Y}'())$ is within a neighborhood of radius $\ast \cdot k e^{k\alpha}$ of its bounded trivial value.

By application of these conclusions in (17) one can find that (20) holds if $\alpha = \chi$ with the left hand side strictly smaller than the right-hand side, provided that $k$ exceeds some constant $k_0$. This result contradicts the assumption that we have equality in (20) if $\alpha = \chi$. Hence, (20) is true for all $\chi$ if $k > k_0$. Details of the last piece of calculation are omitted, but we point out that $\mathbf{C}$ is inside an integral in (17). Integrating its $\ast \varepsilon k \exp(k\alpha)$ maximum deviation from the trivial value yields $\ast \varepsilon \exp(k\chi)$ maximum deviation in $\mathbf{Y}'$ •

*Lemma 2: there exists a positive scalar $\varepsilon_0$ such that (i) and (ii) of Lemma 1 imply that (17) has a unique solution.*

*Proof of Lemma 2:* ODE's with Lipschitz-continuous right-hand sides and given initial condition have unique solutions according to the Picard-Lindelöf theorem (Coddington & Levinson, 1955). We sketch an adaptation of the standard proof of this result to the initial value problem (17).

By introducing the function $\mathbf{\Psi}() = \mathbf{Y}'()$, (17)-(19) can be rewritten as

$$\mathbf{\Psi}(\chi) = \mathbf{A}\left( \chi, \int_0^\chi \mathbf{\Psi}(\beta) d\beta \right)^{-1} \left( \mathbf{F}''(\chi) - \int_0^\chi \mathbf{C}\left( \alpha, \chi, \int_0^\alpha \mathbf{\Psi}(\beta) d\beta, \mathbf{\Psi}(\alpha) \right) \cdot d\alpha \cdot \begin{bmatrix} 1 \\ 1 \end{bmatrix} \right) \tag{25}$$

$$a_{ij} = \left. \frac{\partial}{\partial \chi} G_{ij}\left( \alpha, \chi, \int_0^\alpha \Psi_i(\beta) d\beta \right) \right|_{\alpha = \chi} \tag{26}$$

$$c_{ij} = \Psi_i(\alpha) \cdot \frac{\partial^2}{\partial \chi^2} G_{ij}(\alpha, \chi, \int_0^\alpha \Psi_i(\beta) d\beta) \qquad (27)$$

Assume that the solution $\Psi$ of (25) is known for $\chi \leq \chi_0$ and satisfies (20):

$$\left| \Psi_j(\chi) - (-1)^{j+1} \cos \chi \right| \leq k\varepsilon e^{k\chi} \qquad (28)$$

We wish to prove that there is a unique solution over an additional finite interval $\chi_0 < \chi \leq \chi_1$. Splitting the integrals in (25) at $\chi_0$ yields

$$\Psi(\chi) = \mathbf{A}\left( \chi, \underbrace{\int_0^{\chi_0} \Psi(\beta) d\beta}_{known} + \int_{\chi_0}^{\chi} \Psi(\beta) d\beta \right)^{-1} \cdot \ldots \qquad (29)$$

$$\ldots \left( \underbrace{\mathbf{F}''(\chi) - \int_0^{\chi_0} \mathbf{C}\left(\alpha, \chi, \int_0^\alpha \Psi(\beta) d\beta, \Psi(\alpha)\right) \cdot \begin{bmatrix} 1 \\ 1 \end{bmatrix} d\alpha}_{known} - \int_{\chi_0}^{\chi} \mathbf{C}\left(\alpha, \chi, \int_0^\alpha \Psi(\beta) d\beta, \Psi(\alpha)\right) d\alpha \cdot \begin{bmatrix} 1 \\ 1 \end{bmatrix} \right)$$

The right-hand side of (29) defines a self-map $K$ of the space $S_0$ of vector valued continuous functions $\Psi(\chi)$ over the interval $\chi \in (\chi_0, \chi_1)$. $S_0$ and the metric $d$ induced by the norm

$$\|\Psi(\chi)\| \stackrel{def}{=} \max_{\chi, j} |\Psi_j(\chi)| \qquad (30)$$

form a Banach space. Let $S$ denote the closed subset of $S_0$ determined by (28). The arguments used in the proof of *Lemma 1* imply that for $\varepsilon$ small enough and $k > k_0$, $K$ maps $S$ into itself. The contraction principle implies that if $K$ is a contraction then it has a unique fixed point, corresponding to a unique solution of (29). Repeated application of the above argument yields global existence and uniqueness for $\alpha_1 \leq \chi \leq 1$. Integrating the solution $\Psi$ leads to a unique solution $\mathbf{Y}$ of the original problem.

The only remaining gap in the proof is the contractivity of $K$. $\mathbf{A}$ is Lipschitz in its second variable (cf. (46), details omitted). As $\mathbf{A}$ is nonsingular (Appendix A.2), its inverse is also Lipschitz with some Lipschitz constant $L_{invA}$. Similarly, (27) and some examination of (49) yield that $\mathbf{C}$ is Lipschitz-continuous functional of $\Psi(\chi)$, $\chi \in (\chi_0, \chi_1)$ with a Lipschitz constant $L_C$ (details omitted).

Next, we consider two elements $\Psi^{(1)}$ and $\Psi^{(2)}$ of the set $S$. Then,

$$d\left( \mathbf{A}\left(\chi, \int_0^{\chi_0} \Psi(\beta) d\beta + \int_{\chi_0}^{\chi} \Psi^{(1)}(\beta) d\beta\right)^{-1}, \mathbf{A}\left(\chi, \int_0^{\chi_0} \Psi(\beta) d\beta + \int_{\chi_0}^{\chi} \Psi^{(2)}(\beta) d\beta\right)^{-1} \right) \leq L_{invA}(\chi - \chi_0) d\left(\Psi^{(1)}, \Psi^{(2)}\right) \qquad (31)$$

and

$$d\left( \int_{\chi_0}^{\chi} \mathbf{C}\left(\alpha, \chi, \int_0^{\chi_0} \Psi(\beta) d\beta + \int_{\chi_0}^{\chi} \Psi^{(1)}(\beta) d\beta, \Psi^{(1)}(\alpha)\right) d\alpha, \int_{\chi_0}^{\chi} \mathbf{C}\left(\alpha, \chi, \int_0^{\chi_0} \Psi(\beta) d\beta + \int_{\chi_0}^{\chi} \Psi^{(2)}(\beta) d\beta, \Psi^{(1)}(\alpha)\right) d\alpha \right) \leq \ldots \qquad (32)$$

$$\ldots L_C(\chi - \chi_0) d\left(\Psi^{(1)}, \Psi^{(2)}\right)$$

These inequalities and the boundedness of all terms in the formula of $K$ imply that $d(K(\Psi^{(1)}), K(\Psi^{(2)})) \leq L(\chi - \chi_0) d(\Psi^{(1)}, \Psi^{(2)})$ with some constant $L$ (details omitted). Hence, if $\chi - \chi_0 < L^{-1}$ then $K$ is contractive. •

It follows from the lemmas that

*Theorem 1: for any for any given scalars $0 < \alpha_1 < 1$ and $\Delta > 0$, there exists a positive scalar $\varepsilon_0$ such that (i) and (ii) of Lemma 1 imply that (17) has a unique solution $\mathbf{Y}(\chi)$; $Y_j(\chi) = (-1)^{j+1}\chi$ for $0 \leq \chi \leq \alpha_1$ and $|Y_j(\chi) - (-1)^{j+1}\chi| < \Delta$ for $\alpha_1 < \chi \leq 1$. This solution satisfies (6), (7).*

## 3.2. Characterization of acceptable solutions

The water envelopes $a(\phi)$ and $b(\phi)$ together with a pair of function $Y_j(\phi)$ over the interval $0 \leq \phi \leq 1$ determine the 'upper' ($j=1$) and the 'lower' ($j=2$) half of a unique curve in the *x-y* plane. The curve consists of the points $P_j(\phi) = [(-1)^{j+1} X_i(\phi,\phi,Y_j(\phi)); Y_j(\phi)]$. Rotation of this curve about the *y* axis generates a unique object with cylindrical symmetry. Below we state a sufficient condition under which the object is simple. This condition is satisfied by the nontrivial solutions predicted by *Theorem 1*.

*Lemma 3:* If

$$\left. \begin{array}{ll} \text{either} & (-1)^{j+1} Y_j(\phi) > \delta \\ \text{or} & (-1)^{j+1} X_j(\phi,\phi,Y_j(\phi)) > \delta \end{array} \right\} \text{ for any } 0 < \phi < 1 \qquad (33)$$

with

$$\delta = \max \left\{ \begin{array}{l} \max_\phi |a(\phi)| \\ \max_\phi |b(\phi)| \end{array} \right\} \qquad (34)$$

*then the object is a simple topological ball.*

*Proof of Lemma 3:*
The condition of the lemma means that one can draw a square of size $2\delta \times 2\delta$ about the origin of the *x-y* plane such that the water envelope is inside the square while the contour curve is outside (Fig. 2). Outside the square, the upper-right quarter of the *x-y* plane is covered with the non-intersecting lines $L(\phi)$ each containing a point $P_1(\phi)$. The lower-right quarter contains the points $P_2(\phi)$ each one lying on the mirror image of line $L(\phi)$ about the *y* axis. Thus, all points $P_i(\phi)$ for $\phi<1$ are in the right half-plane, separated from the *y* axis. Furthermore, two points of the contour curve corresponding to different values of $\phi$ or different values of *i* lie on different lines, hence they may not coincide. Altogether we have found that the *contour curve does not touch the y axis* (except at the endpoints: $\phi=1$), and *it is not self-intersecting* (or self-touching). Rotating such curves generates topological balls.

Due to the cylindrical symmetry of the object, being simple is equivalent of requiring that $L^*(\phi)$ is a connected line segment for every $\phi$ (rather than the union of multiple segments). $L^*(0)$ is connected, hence, the object is simple iff by varying $\phi$, the topology of $L^*(\phi)$ does not change. A topological change of $L^*$ occurs at $\phi$ if the contour curve touches $L(\phi)$ at $P_1(\phi)$ or the mirror image of $L(\phi)$ at $P_2(\phi)$ *without crossing the line*. Nevertheless this situation is impossible because, as already mentioned, the set of lines $L(\phi)$ underlying the points $P_i(\phi)$ is free of intersections outside the square.•

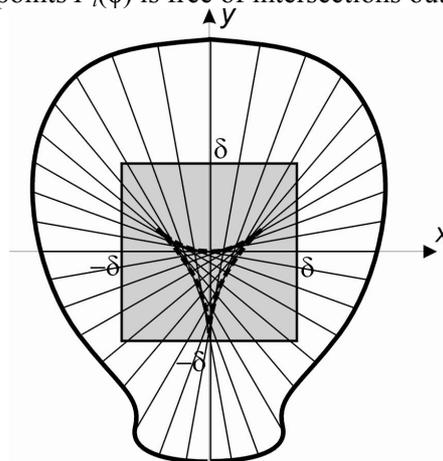

**Fig. 2: illustration of *Lemma 3*: if the water envelope is inside the grey square, and the contour curve is outside, then the object is a simple topological ball.**

## 3.3. Water envelopes and numerical examples

A convenient way to find a suitable water envelope is to pick a $C^1$ function $\rho(\Phi)$ with a bounded but possibly discontinuous second derivative representing the signed radius of curvature of the water envelope at tangent angle $\Phi$. Then,

$$A(\Phi) = \int_{\Phi}^{\pi/2} \rho(\Theta)\cos\Theta\, d\Theta + c_A \tag{35}$$

$$B(\Phi) = \int_{\Phi}^{\pi/2} \rho(\Theta)\sin\Theta\, d\Theta + c_B \tag{36}$$

The symmetry of the problem dictates that
- A: $A(\pi/2)=0$, hence $c_A=0$;
- B: variations of $c_B$ result in translated copies of the same envelope, i.e. we can choose $c_B=0$.
- C: $A(0)=0$, which is a constraint on $\rho(\Phi)$ by **(35)**;
- D: $\rho(\Phi)$ is $\pi$-periodic and even;
- E: $\rho(\Phi-\pi/2)$ is odd, implying $\rho(\pi/2)=0$.

In the variables $\alpha$ and $\phi$, **(35)** and **(36)** become

$$a(\phi) = \int_{\phi}^{1} \rho(\arcsin\alpha)\, d\alpha \tag{37}$$

$$b(\phi) = \int_{\phi}^{1} \rho(\arcsin\alpha)\frac{\alpha}{\sqrt{1-\alpha^2}}\, d\alpha \tag{38}$$

The $\arcsin\alpha$ function has a square-root singularity at $\alpha=1$. According to observation E, $\rho(\arcsin\alpha) \approx \textit{constant} \cdot (\pi/2-\alpha)^{1/2}$ near $\phi=1$. This singularity is cancelled by a $(\pi/2-\alpha)^{-1/2}$ term in **(38)**, thus $b(\phi)$ becomes $C^2$ with a bounded third derivative. At the same time, $a(\phi) \approx \textit{constant} \cdot (1-\phi^2)^{3/2}$ near $\phi=1$, which means that $a(\phi)$ is singular at $\phi=1$, but $\tilde{a}(\phi) = a(\phi)(1-\phi^2)^{1/2}$ has a bounded third derivative. The smoothness of $b$ and $\tilde{a}$ mean that any function $\rho(\Phi)$ multiplied by a sufficiently small constant meets condition (ii) of *Lemma 1*.

Two examples fulfilling the above requirements are

$$\rho(\Phi) = c \cdot \cos((2n+1)\Phi) \qquad n = 1,2,3... \tag{39}$$

$$\rho(\Phi) = c \cdot \begin{cases} 0 & \text{if} \quad \phi \leq \pi/4 \\ \sin^2 4\Phi - \dfrac{85}{84}\sin^3(4\Phi) & \text{if} \quad \pi/4 < \phi \leq \pi/2 \end{cases} \tag{40}$$

where $c$ is an arbitrary constant; the number 85/84 is determined by Observation C. The second example obeys condition (i) of the lemma, hence this envelope generates a nontrivial solution by *Theorem 1* if its unspecified constant is small enough, see also Fig. 3A.

The first example does not meet condition (i) nevertheless the solution appears to exist and to be unique in this case, too (Fig. 3B-D). Indeed, condition (i) is probably unnecessary for *Lemma 1*, but it simplifies the proof (see Appendix A.3). Additionally, condition (i) has a central role in the proof of *Lemma 2*. Nevertheless, existence (and uniqueness) of the solution might be provable with a different approach without condition (i).

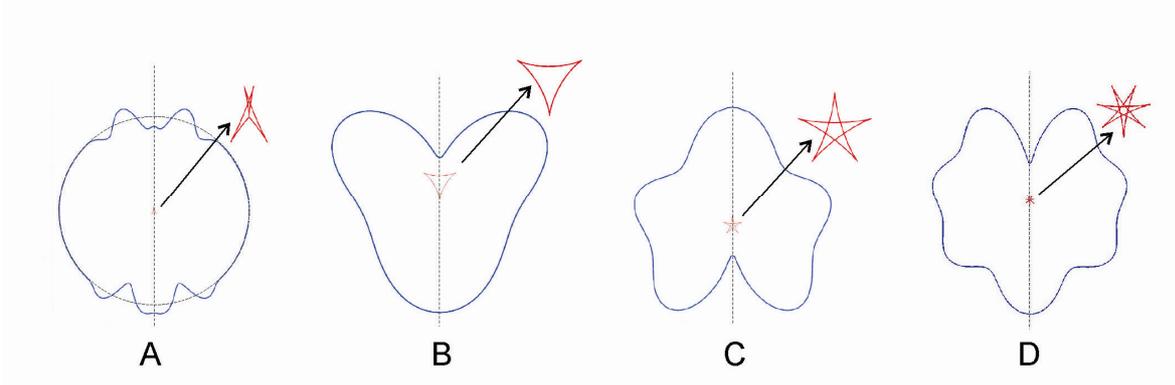

**Figure 3:** Numerically determined contour curves and water envelopes of some neutrally floating shapes with cylindrical symmetry. A: water envelope (34) with $c=0.5$; B-D: water envelope (33) with $n=1,2,3$ and $c=0.5; 0.5; 0.4$. In all cases, $Y_1^{'}(0)=-Y_2^{'}(0)=1$.

## 4. Discussion

This paper is concerned with the proof of existence and the construction of neutrally floating, simple objects of density 1/2 (other than the sphere) in three dimensions. As we show, there are many solutions even among bodies with cylindrical symmetry. Our study leaves many open questions, including the necessity of condition (i) in *Theorem 1*, or the existence of solutions for densities other than 1/2.

The present discussion of the Floating Body Problem concentrates on gravitational (and buoyancy) forces, and excludes any other forces acting on the object. A different approach has been taken by R. Finn and coworkers (Finn, 2009; Finn & Sloss, 2009), see also Gutkin (2010), who studied the same question for objects floating in gravity-free environment under the effect of capillary forces. In this approach, the contact angle of the object and the liquid is a free parameter analogous to density in the presence of gravity. The two-dimensional capillary floating problem admits nontrivial solutions similarly to the Archimedean version. In three dimensions, only a special nonexistence result has been published: spheres are the only objects, which can float in any orientation in such a way that the capillary forces *preserve the flatness of the liquid surface*. In most cases, a macroscopically flat liquid surface typically becomes distorted in a small neighborhood of a floating object to minimize the surface energy of the system. This more general situation seems to be unexplored.

While gravity-free floating may appear as a weird setting at first sight, it is physically as relevant as the Archimedean approach. Physical systems under terrestrial conditions are inevitably subject to both gravity and capillary forces. The relative strengths of the two forces are determined by the dimensionless Eötvös- (or Bond-) number of the system (R. Finn, oral communication). Small-scale objects have low Eötvös number (indicating the dominance of capillary effects), whereas upscaling an object increases the Eötvös number. For example, the Eötvös number of a ball of density ½ and radius $r$ floating in water is approximately $(r/4\text{mm})^2$. Thus, the dominance of each of the two effects can be realized in a physical experiment. Additionally, there exists a generalized – and completely unexplored - version of Ulam's problem, which seeks neutrally floating objects under dual influence of gravity and surface tension for given density, contact angle and Eötvös number.

## A Appendix

The appendix contains several technical results needed for *Lemma 1*.

### A.1 Deviation of $a_{ij}$ from its trivial value

$a_{ij}$ is given by (18) as a partial derivative of $G_{ij}$, which is defined by an improper integral (13). We define a new function $Q_i$

$$Q_j(\alpha,\phi,Y_j(\alpha)) \stackrel{def}{=} \begin{cases} Z_j(\alpha,\phi,Y_j(\alpha))^2\left(1-\dfrac{\alpha}{\phi}\right)^{-1} & \text{if } \alpha<\phi \\ \lim\limits_{\beta\to\phi^-} Q_i(\beta,\phi,Y_j(\beta)) & \text{if } \alpha=\phi \end{cases} \qquad (41)$$

which proves useful later, in Appendix A.3. This definition is motivated by the square-root type singularity of $Z_i$ at $\alpha=\phi$, which implies that $Q_i$ is bounded and strictly positive. With the new function and equations (6)-(8),(13) we obtain

$$G_{ij}(\alpha,\chi,Y_j(\alpha)) = \int_\alpha^\chi (\chi-\phi)^{-1/2}\left(1-\dfrac{\alpha}{\phi}\right)^{1/2} \sqrt{Q_j(\alpha,\phi,Y_j(\alpha))}\,(Y_j(\alpha)-b(\phi))^j\,d\phi \qquad (42)$$

The variable $\phi$ of integration is changed to $\Gamma=(\chi-\phi)^{1/2}\cdot(\chi-\alpha)^{-1/2}$:

$$G_{ij}(\alpha,\chi,Y_j(\alpha)) = 2(\chi-\alpha)\int_0^1 \left(\chi+\alpha\dfrac{\Gamma^2}{1-\Gamma^2}\right)^{-1/2} Q_j^{1/2}(\alpha,\chi-\Gamma^2(\chi-\alpha),Y_j(\alpha))\,(Y_j(\alpha)-b(\chi-\Gamma^2(\chi-\alpha)))^j\,d\Gamma \qquad (43)$$

This form of $G_{ij}$ is a proper integral, and also free of terms diverging to infinity at $\alpha=\chi$. $a_{ij}$ can now be calculated from (18) and (43) by using the Leibniz rule, then by plugging $\alpha=\chi$, and finally by evaluating a simple integral:

$$a_{ij}(\chi,Y_j(\chi)) = \dfrac{\pi}{2}\chi^{-1/2} Q_j^{1/2}(\chi,\chi,Y_j(\chi))\,(Y_j(\chi)-b(\chi))^i \qquad (44)$$

The formula above contains $Q_j(\chi,\chi,Y_j(\chi))$, which can be expressed as a function of $X_j()$:

$$Q_j(\chi,\chi,Y_j(\chi)) = \lim_{\alpha\to\chi^-} \dfrac{X_j^2(\alpha,\alpha,Y_j(\alpha))-X_j^2(\alpha,\chi,Y_j(\alpha))}{1-\alpha/\chi} = \ldots \qquad (45)$$

$$-\chi \lim_{\alpha\to\chi} \dfrac{X_j^2(\alpha,\chi,Y_j(\alpha))-X_j^2(\alpha,\alpha,Y_j(\alpha))}{\chi-\alpha} = \ldots$$

$$-\chi \dfrac{\partial}{\partial\chi} X_j^2(\alpha,\chi,Y_j(\alpha))\bigg|_{\alpha=\chi}$$

We plug this equation into (44) and use (3) to express $a_{ij}$ explicitly as a function of $Y_j$:

$$a_{ij}(\chi,Y_j(\chi)) = \dfrac{\pi}{\sqrt{2}}\left(-\chi\dfrac{\partial X_j^2(\alpha,\chi,Y_j(\alpha))}{\partial\chi}\bigg|_{\alpha=\chi}\right)^{1/2}(Y_j(\chi)-b(\chi))^j = \qquad (46)$$

$$= \dfrac{\pi}{2}\sqrt{-\dfrac{\partial}{\partial\chi}\left[(\tilde{a}(\chi)\chi+Y_j(\alpha)-b(\chi))^2(\chi^{-2}-1)\right]\bigg|_{\alpha=\chi}}\cdot(Y_j(\chi)-b(\chi))^j$$

Eq. (21) and point (ii) of *Lemma 1* can be expressed as: $Y_j(\chi)\in(-1)^{j+1}\chi\pm\varepsilon\exp(k\chi)$ and $\tilde{a}(\chi),b(\chi),\tilde{a}'(\chi)$, $b'(\chi)\in\pm\varepsilon\in\pm\varepsilon\exp(k\chi)$. Plugging these into (46) together with the inequality $\chi\geq\alpha_1$; replacing higher order terms of $\varepsilon\exp(k\chi)$ by a small constant time $\varepsilon\exp(k\chi)$; and noting that the term under the square-root sign has a strictly positive lower bound lead to the final expression

$$a_{ij}(\chi,Y_j(\chi))\in(-1)^{i(j+1)}\dfrac{\pi}{\sqrt{2}}\chi^{i-1/2}\pm k_1\varepsilon e^{k\chi} \qquad (47)$$

with some positive constant $k_1$ not specified for brevity. This is the result we had to prove.

### A.2 An upper bound of $|(a_{11}a_{22}-a_{12}a_{21})^{-1}|$

An approximation of $a_{ij}$ with $*\varepsilon\exp(k\chi)$ uncertainty has been given by (47). This formula yields

$$a_{11}a_{22}-a_{12}a_{21}\in\pi^2\chi^2\pm *\varepsilon\exp(k\chi) \qquad (48)$$

Since $\chi>\alpha_1$, we have found a positive lower bound of $a_{11}a_{22}-a_{12}a_{21}$.

## A.3 The second derivative of $G_{ij}$

This section is devoted to the proof of equations (22) and (23). The second derivative of $G_{ij}$ is calculated from (43) by successive applications of the Leibniz rule. The result (calculated by Maple software and not shown) can be written in the form

$$\frac{\partial^2 G_{ij}(\alpha,\chi,Y_j(\alpha))}{\partial \chi^2} = \int_0^1 \sum_k (W_{1k} \cdot W_{2k} \cdot ...)d\Gamma \qquad (49)$$

where $W_{lk}$ are functions of $i,j,\alpha,\chi,\Gamma,Y_j(\alpha)$, $a()$ and $b()$. Specifically, they include constants and eight non-constant terms listed in Table 1.

Table 1: a full list of terms emerging in the second partial derivative of $G_{ij}$.

| Number | Term |
|---|---|
| 1 | $\left(\chi + \alpha \dfrac{\Gamma^2}{1-\Gamma^2}\right)^{-1/2}$ |
| 2 | $Y_j(\alpha) - b(\chi - \Gamma^2(\chi - \alpha))$ |
| 3 | $b'(\chi - \Gamma^2(\chi - \alpha))$ |
| 4 | $b''(\chi - \Gamma^2(\chi - \alpha))$ |
| 5 | $Q_j(\alpha,\phi)^{1/2}$  where $\phi = \chi - \Gamma^2(\chi - \alpha)$ |
| 6 | $Q_j(\alpha,\phi)^{-1/2}$  where $\phi = \chi - \Gamma^2(\chi - \alpha)$ |
| 7 | $(1-\Gamma^2)\left.\dfrac{\partial Q_j(\alpha,\phi)}{\partial \phi}\right|_{\phi = \chi - \Gamma^2(\chi-\alpha)}$ |
| 8 | $(1-\Gamma^2)^2 \left.\dfrac{\partial^2 Q_j(\alpha,\phi)}{\partial \phi^2}\right|_{\phi = \chi - \Gamma^2(\chi-\alpha)}$ |

As we show below, each term has a bounded absolute value, and its deviation from its trivial value is at most *constant*·$\varepsilon\exp(k\alpha)$.

- **Term 1** is not bigger than $\chi^{-1/2}$, hence it is bounded from above by the constant $\alpha_1^{-1/2}$; this term is not affected by perturbations of the water envelope, since it does not depend on any of the functions $Y_j$, $a$ or $b$.
- **Term 2** is $Y_j(\alpha) - b(\chi - \Gamma^2(\chi-\alpha)) \in (-1)^{j+1}\alpha \pm \varepsilon e^{k\alpha} \pm \varepsilon \in (-1)^{j+1}\alpha \pm 2\varepsilon e^{k\alpha}$
- **Term 3 and 4** are $\in 0 \pm \varepsilon \in 0 \pm \varepsilon e^{k\alpha}$
- **Term 5 and 6**: it is enough to show that $Q_j(...)$ itself has an absolute value bounded from above *and below* by positive bounds, and that it is affected by at most a constant times $\varepsilon\exp(k\alpha)$ by the perturbation. By using (3), (4) and **(41)**, expanding the nominator and ordering its terms into pairs (marked by square brackets), $Q_j$ can be expressed as

$$Q_j(\alpha,\phi,Y_j(\alpha)) = \frac{(\tilde{a}(\alpha) + (Y_j(\alpha)-b(\alpha))\alpha^{-1})^2(1-\alpha^2) - (\tilde{a}(\phi)+(Y_j(\alpha)-b(\phi))\phi^{-1})^2(1-\phi^2)}{1-\alpha/\phi} = ... \qquad (50)$$

$$= \phi\frac{[\tilde{a}(\alpha)^2 - \tilde{a}(\phi)^2] - [\alpha^2\tilde{a}(\alpha)^2 - \phi^2\tilde{a}(\phi)^2] + Y_i^2(\alpha)[\alpha^{-2} - \phi^{-2}] + etc.}{\phi - \alpha} = ...$$

$$= \frac{\left[\overbrace{(\tilde{a}(\alpha)-\tilde{a}(\phi))}^{\in\pm\varepsilon(\phi-\alpha)}\overbrace{(\tilde{a}(\alpha)+\tilde{a}(\phi))}^{\in\pm 2\varepsilon}\right]\phi - \left[\overbrace{(\alpha\tilde{a}(\alpha)-\phi\tilde{a}(\phi))}^{\in\pm 2\varepsilon(\phi-\alpha)}\overbrace{(\alpha\tilde{a}(\alpha)+\phi\tilde{a}(\phi))}^{\in\pm 2\varepsilon}\right]\phi + \left[\overbrace{\phi Y_j^2(\alpha)(\alpha^{-2} - \phi^{-2})}^{\in\left(1+\frac{\alpha}{\phi}\right)(\phi-\alpha)\pm\varepsilon e^{k\alpha}2\alpha_1^{-3}(\phi-\alpha)}\right] + etc. ...}{\phi - \alpha}$$

Only a few terms of the nominator are shown. Each square bracket in the nominator can be expressed as $\ast \cdot (\phi-\alpha)$ in order to cancel the denominator. This step is straightforward in some cases, but less so in others. For example, in the case of the first two square brackets, we use that $|\tilde{a}(\alpha)|, |\tilde{a}'(\alpha)| < \varepsilon$; at the third one, we exploit that $Y_j(\alpha)=(-1)^{j+1}\alpha$ if $\alpha \leq \alpha_1$; whereas (21) holds and $\alpha^{-2} - \phi^{-2} \in \pm 2\alpha_1^{-3}(\phi-\alpha)$ if $\alpha > \alpha_1$. This calculation leads to

$$Q_j \in 1 + \alpha/\phi \pm \ast \cdot \varepsilon \exp(k\alpha). \qquad (51)$$

- **Term 7:** The trivial value of $Q_j$ is given by (51). This is used to determine the trivial value of term 7, which is bounded because

$$\left(1-\Gamma^2\right) \left. \frac{\partial Q_j(\alpha, \phi, Y_j(\alpha))}{\partial \phi} \right|_{\phi = \chi - \Gamma^2(\chi-\alpha)} = \frac{\phi-\alpha}{\chi-\alpha} \cdot \left(-\frac{\alpha}{\phi^2}\right) = \ldots \qquad (52)$$

$$= \begin{cases} \dfrac{\phi-\alpha}{\chi-\alpha} \dfrac{\alpha}{\phi} \phi^{-1} \leq 1 \cdot 1 \cdot \left(\dfrac{\alpha_1}{2}\right)^{-1} & \text{if } \alpha \geq \chi/2 \\ \dfrac{\phi-\alpha}{\phi} \dfrac{\alpha}{\phi} (\chi-\alpha)^{-1} \leq 1 \cdot 1 \cdot \left(\dfrac{\alpha_1}{2}\right)^{-1} \leq & \text{if } \alpha \leq \chi/2 \end{cases}$$

The deviation of term 7 from its trivial value can be obtained analogously to the calculations of term 5 and 6 by exploiting that the first *and second* derivatives of $\tilde{a}$ and $b$ are $\in \pm\varepsilon$. These are not shown for brevity.

- **Term 8:** the trivial value can be investigated using a formula analogous to (52):

$$\left(1-\Gamma^2\right)^2 \cdot \left. \frac{\partial Q_j(\alpha, \phi, Y_j(\alpha))}{\partial \phi} \right|_{\phi = \chi - \Gamma^2(\chi-\alpha)} = \left(\frac{\phi-\alpha}{\chi-\alpha}\right)^2 \frac{2\alpha}{\phi^3} = \ldots \qquad \mathbf{(53)}$$

$$= \begin{cases} 2\left(\dfrac{\phi-\alpha}{\chi-\alpha}\right)^2 \dfrac{\alpha}{\phi} \phi^{-2} \leq 2 \cdot 1 \cdot 1 \cdot \left(\dfrac{\alpha_1}{2}\right)^{-2} & \text{if } \alpha \geq \chi/2 \\ 2\left(\dfrac{\phi-\alpha}{\phi}\right)^2 \dfrac{\alpha}{\phi} (\chi-\alpha)^{-2} \leq 2 \cdot 1 \cdot 1 \cdot \left(\dfrac{\alpha_1}{2}\right)^{-2} & \text{if } \alpha \leq \chi/2 \end{cases}$$

As before, explicit bounds of the nontrivial value are not shown. They can be obtained by exploiting that the derivatives of $\tilde{a}$ and $b$ up to third order are $\in \pm\varepsilon$.

Both of these properties are inherited by the second derivative of $G_{ij}$ according to eq. (49), yielding (22) and (23).